\documentclass[twocolumn,superscriptaddress,showpacs,preprintnumbers,amsmath,amssymb]{revtex4}
\usepackage{bm,graphicx}
\newcommand{\bb}{\begin{equation}}
\newcommand{\en}{\end{equation}}

\begin{document}

\title{Non-equilibrium fluctuations and mechanochemical couplings of a molecular motor}

\author{A.W.C. Lau}
\affiliation{Department of Physics, Florida Atlantic University,
777 Glade Rd., Boca Raton, FL 33431}
\author{D. Lacoste}
\affiliation{Laboratoire de Physico-Chimie Th\'eorique, UMR 7083, ESPCI,
10 rue Vauquelin, 75231 Paris cedex 05, France}
\author{K. Mallick}
\affiliation{Service de Physique Th\'eorique, CEA Saclay, 91191 Gif, France}

\begin{abstract}
We investigate theoretically the violations of Einstein and Onsager relations,
and the efficiency for a single processive motor operating far from equilibrium using an extension of the
two-state model introduced by Kafri {\em et al.} [Biophys.\ J.\ {\bf 86}, 3373 (2004)].
With the aid of the Fluctuation Theorem, we analyze the general features of these violations and
this efficiency and link them to mechanochemical couplings of motors.  In particular,
an analysis of the experimental data of kinesin using our framework leads to interesting predictions
that may serve as a guide for future experiments.
\end{abstract}

\date{\today}
\maketitle


Motor proteins are nano-machines that convert chemical energy into mechanical
work and motion \cite{howard}.  Important examples include kinesin, myosin,
and RNA polymerase.  Despite a number of theoretical
models \cite{armand1,parmeggiani,kolomeisky,nelson,others},
understanding the mechanochemical transduction mechanisms behind these
motors remains a significant challenge \cite{motorreview}.
Recent advances in experimental techniques \cite{block,coppin} to probe the
fluctuations of single motors provide ways to gain insight into their kinetic pathways \cite{schnitzer}.
However, a general description for fluctuations of systems driven out of equilibrium,
and in particular of motors, is still lacking.
Recently, the Fluctuation Theorem (FT) \cite{FT,gallavotti,lebowitz}
has emerged as a promising framework to characterize fluctuations in far-from-equilibrium regimes where
Einstein and Onsager relations no longer hold \cite{gallavotti}.  In a nutshell, FT states
that the probability distribution for the entropy production rate
obeys a symmetry relation, and it has been verified
in a number of beautiful experiments on biopolymers and colloidal systems \cite{ritort}.
In this Letter, we demonstrate that FT provides a natural framework in which thermodynamic constraints
can be imposed on the operation of nano-machines far from equilibrium.

Specifically, we study a generalization of the two-state model of
motors introduced in Ref.\ \cite{nelson}.  Although similar models
have been investigated with known exact results
\cite{kolomeisky,nelson}, we reformulate the model to include an
important variable, namely the number of ATP consumed, and
construct a thermodynamic framework.  Our framework allows us to
characterize the ATP consumption rate of a motor, its run length,
and its thermodynamic efficiency. Additionally, we show that our
model obeys FT \cite{lebowitz}. While there have been a few recent
studies proving FT for motors \cite{qian,seifert,gaspard}, we
further investigate the physical implications of FT here.  In
particular, we quantify the violations of Einstein and Onsager
relations, respectively, by four temperature-like parameters,
$T_{ij}$, and by the difference of the mechanochemical coupling
coefficients, $\Delta \lambda$, and we explore the behaviors of
$T_{ij}$ and $\Delta \lambda$, as well as the motor efficiency, as
functions of generalized forces with the aid of FT.  Our main
results are (i) one of the Einstein relations holds near stalling,
(ii) the degree by which the Onsager symmetry is broken ($\Delta
\lambda \neq 0$) is largely determined by the underlying asymmetry
of the substrate, (iii) only two ``effective" temperatures
characterize the fluctuations of tightly coupled motors, and (iv)
kinesin's maximum efficiency and its maximum violation of Onsager
symmetry occur roughly at the same energy scale, corresponding to
that of an ATP hydrolysis ($\sim 20\,k_B T$).

\begin{figure}
\includegraphics[height=1.45in]{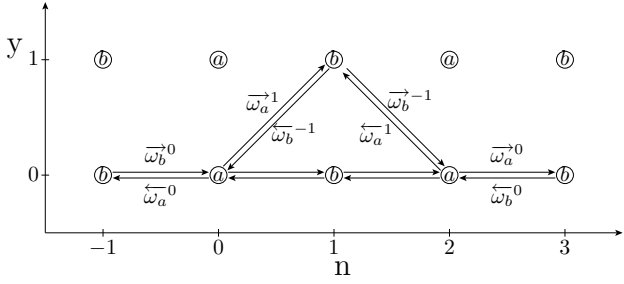}
\caption{A schematic of the rates of the two-state model for a
molecular motor moving on a linear lattice with $y$ number of ATP
consumed. The even and odd sites are denoted by $a$ and $b$,
respectively. In the case of two headed kinesin, site $a$
represents a state where both heads are bound to the filament,
whereas site $b$ represents a state with only one head bound.}
\label{fig:sketch}
\end{figure}

As a result of conformational changes powered by hydrolysis of
ATP, a linear processive motor, like kinesin, moves along a
one-dimensional substrate (microtubules).  Its state may be
characterized by two variables: its position and the number of ATP
consumed. To model its dynamics, we consider a linear discrete
lattice, where the motor ``hops" from one site to neighboring
sites, either consuming or producing ATP (see
Fig.~\ref{fig:sketch}). The position is denoted by $x = n d$,
where $2 d \approx 8 \, \mbox{nm}$ is the step size for kinesin.
The even sites (denoted by $a$) are the low-energy state of the
motor, whereas the odd sites (denoted by $b$) are its high-energy
state;  their energy difference is $ \Delta E \equiv k_B T
\epsilon$, where $k_B$ is the Boltzmann constant and $T$ is the
temperature.

Because of the periodicity of the filament, all the even ($a$)
sites and all the odd ($b$) sites are equivalent. The dynamics is
governed by a master equation for the probability, $P_i(n,y,t)$,
that the motor, at time $t$, has consumed $y$ units of ATP and is at site
$i$ $(=a,b)$ with position $n$: \begin{eqnarray}
\partial_t P_i(n,y,t)  = -  \left( \overleftarrow{\omega}_i + \overrightarrow{\omega}_i
\right) P_{i}(n,y,t)+
\phantom{\overleftarrow{\omega}_{n+1}^{\,l}ppppppp}\nonumber \\
\sum_{l = -1,0,1} \left [\,
\overleftarrow{\omega}_{j}^{\,l}\,P_{j}(n+1,y-l,t)+
\overrightarrow{\omega}_{j}^{\,l}\,P_{j}(n-1,y-l,t) \,\right ],\nonumber
\end{eqnarray}
with $i \neq j$, where $\overleftarrow{\omega}_j \equiv \sum_l
\overleftarrow{\omega}_{j}^{\,l}$ and $\overrightarrow{\omega}_j
\equiv \sum_l \overrightarrow{\omega}_{j}^{\,l}$. Denoted by
$\overleftarrow{\omega}_{j}^{\,l}$ and
$\overrightarrow{\omega}_{j}^{\,l}$ are the transition probability
per unit time for the motor, with $l\,(= -1,0,1)$ ATP molecules
consumed, to jump from site $j$ to a neighboring site to the left
or to the right, respectively.

The transition rates can be constructed by considering the
kinetics of the transitions between the two states $\mbox{M}_a$
and $\mbox{M}_b$ of the motor \cite{parmeggiani}. We assume two
different chemical pathways: ($\alpha$) $\mbox{M}_a + \mbox{ATP}
\rightleftharpoons \mbox{M}_b + \mbox{ADP} + \mbox{P}$, ($\beta$)
$\mbox{M}_a  \rightleftharpoons  \mbox{M}_b$. The $\alpha$-pathway
represents the transition of the motor accompanied by ATP
hydrolysis and the $\beta$-pathway represents the transition
driven by thermal activation. It is straightforward to generalize the
model with more chemical pathways, but here we focus
only on these two, for which
$\overleftarrow{\omega_b}^{1}=\overrightarrow{\omega_a}^{-1}
=\overleftarrow{\omega_a}^{-1}=\overrightarrow{\omega_b}^{1}=0$.
Following Ref.\ \cite{kolomeisky}, the transition rates in the
presence of an external force $F_e$ are changed according to
$\overleftarrow{\omega}_{i}^{\,l}(F_e) =
\overleftarrow{\omega}_{i}^{\,l}(0)\,e^{-\theta^{-}_i f}$ and
$\overrightarrow{\omega}_{i}^{\,l}(F_e) =
\overrightarrow{\omega}_{i}^{\,l}(0)\,e^{+\theta^{+}_i f}$, where
$f \equiv F_{e} d /( k_B T)$ and $\theta^{\pm}_i$ are the load
distribution factors \cite{kolomeisky}. These load distribution factors
take into account that the external force may not distribute
uniformly  among different transitions. After one period, the
work done by $F_e$ on the motor is $-F_e 2d$, implying that
$\theta^{+}_a+\theta^{-}_b + \theta^{-}_a+\theta^{+}_b\
=2$ \cite{kolomeisky}. Thus, we may write the non-zero rates as:\bb
\begin{array}{ll}
\overleftarrow{\omega_b}^{-1} = \alpha\, e^{-\theta^{-}_b f}, & \overleftarrow{\omega_b}^0 =  \omega\,e^{-\theta^{-}_b f},  \\
\overrightarrow{\omega_a}^1   =  \alpha\, e^{ -\epsilon + \Delta \mu + \theta^{+}_a f }, & \overrightarrow{\omega_a}^0 = \omega\,e^{-\epsilon + \theta^{+}_a f},    \\
\overleftarrow{\omega_a}^1  =  \alpha' e^{ -\epsilon + \Delta \mu - \theta^{-}_a f}, & \overleftarrow{\omega_a}^0  = \omega'\,e^{-\epsilon- \theta^{-}_a f},\\
\overrightarrow{\omega_b}^{-1} = \alpha' e^{\theta^{+}_b f}, & \overrightarrow{\omega_b}^0   =  \omega'\,e^{\theta^{+}_b f},
\end{array}
\label{rates} \en where $\alpha$ and $\alpha'$, and $\omega$ and
$\omega'$ are the bare rates for the two distinct transitions for
the pathways and $\Delta \widetilde{\mu} \equiv k_B T \Delta \mu$
is the chemical potential difference \cite{parmeggiani}. The
underlying asymmetry of the substrate dictates that $\alpha \neq
\alpha'$ and $\omega \neq \omega'$ as required for directional
motion \cite{nelson}.

We find that the rates in Eq.\ (\ref{rates}) satisfy four
generalized detailed balance conditions:
\begin{eqnarray} \overrightarrow{\omega_b}^{-l}
P_b^{\mbox{\scriptsize eq}} & = & \overleftarrow{\omega_a}^{l}
P_a^{\mbox{\scriptsize eq}}\,e^{+(\,\theta^{-}_a+\theta^{+}_b\,)f
- \Delta \mu\,l},
\label{Generalized db1}\\
\overleftarrow{\omega_b}^{-l}\,P_b^{\mbox{\scriptsize eq}} & = &
\overrightarrow{\omega_a}^{l} P_a^{\mbox{\scriptsize eq}}\,e^{- (\,\theta^{+}_a+\theta^{-}_b\,)f - \Delta \mu\,l},
\label{Generalized db2}
\end{eqnarray}
for $l=0,1$, where $P_a^{\mbox{\scriptsize eq}} =
1/(1+e^{-\epsilon})$ and $P_b^{\mbox{\scriptsize eq}} =
e^{-\epsilon}/(1+e^{-\epsilon})$ are the equilibrium probabilities
corresponding to $f=0$ and $\Delta \mu =0$. We note that these relations,
Eqs.~(\ref{Generalized db1}) and (\ref{Generalized db2}), while
valid arbitrary far from equilibrium, still refer to the equilibrium state via
the probabilities $P_i^{eq}$. We show below that these relations
lead to a FT \cite{lebowitz}. Introducing the generating
functions: $F_i(z_1,z_2,t) \equiv \sum_{y} \sum_{n}
e^{-z_1 n - z_2 y} P_{i}(n,y,t),$ whose time evolution is governed
by: $\partial_t F_i= {\cal M}_{ij}\,F_j$, where ${\cal M}[z_1,
z_2]$ is a $2\times 2$ matrix that can be obtained from the master
equation above, we find $ \left \langle\,e^{-z_1 n - z_2
y}\,\right \rangle = \sum_{i} F_i(z_1,z_2,t) \sim \exp \left(
\vartheta\,t \right)$, for $t \rightarrow \infty$, where $
\vartheta \equiv \vartheta[z_1,z_2]$ is the largest eigenvalue of
${\cal M}$. Using Eqs.\ (\ref{Generalized db1}) and
(\ref{Generalized db2}), it can be shown that ${\cal M}$ and ${\cal M}^{\dag}$ are related by
a similarity transformation: ${\cal M}^{\dag}[\,f  - z_1 ,
\Delta \mu - z_2 ]=  {\cal Q}\,{\cal M}[z_1 , z_2]\,{\cal Q}^{-1}$,
where ${\cal M}^{\dag}$ is the adjoint of ${\cal M}$ and ${\cal
Q}$ is a diagonal matrix. This similarity relation implies that\bb
\vartheta[z_1,z_2]= \vartheta[\,f -z_1,\Delta \mu - z_2],
\label{FT} \en which is one form of FT.

\begin{figure}
\includegraphics[height=1.9in]{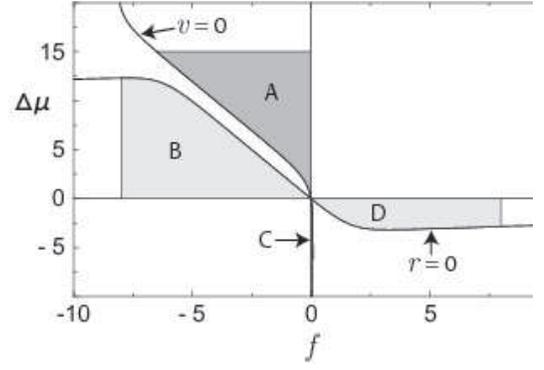}
\caption{Four modes of operation of a molecular motor, as
delimited by $\hat{v}=0$ and $r=0$ \cite{armand1}.  The lines are generated with parameters
that we have extracted from fitting the data for kinesin in Ref.~\cite{block} to our model.
(The best-fit values for the parameters are listed at the end of the text.)
In Region A, where $r \Delta \mu >0$ and $f \hat{v}<0$,
the motor uses chemical energy of ATP to perform mechanical work.  In Region B, where
$r \Delta \mu <0$ and $f \hat{v}>0$, the motor produces ATP from mechanical work.
In Region C, where $r \Delta \mu >0$ and $f \hat{v} <0$, the motor uses ADP to
perform mechanical work. In Region D, where $r \Delta \mu <0$ and $f \hat{v}>0$,
the motor produces ADP from mechanical work.}
\label{fig:modes_operation}
\end{figure}

Now, we proceed to discuss the physical consequences of FT. The
eigenvalue, $\vartheta$, contains all the steady-state properties
of the motor. In particular, the average (normalized) velocity,
$\hat{v} = v/d$, and the average ATP consumption rate, $r$, are,
by definition, given by $\hat{v} = - {
\partial_{z_1}\vartheta }[\,0,0]$ and $r= -{ \partial_{z_2}
\vartheta }[\,0,0]$, respectively \cite{notation}. From the conditions: $\hat{v}=0$
and $r=0$, we can construct a full operation diagram of a motor, as shown in Fig.\
\ref{fig:modes_operation} for the case of kinesin. The curves $\hat{v}=0$ and
$r=0$ define implicitly $f = f_{\mbox{\scriptsize
st}}(\Delta \mu)$ (the stalling force) and $\Delta \mu = \Delta
\mu_{\mbox{\scriptsize st}}(f)$, respectively. It is interesting
to note that the large asymmetry between regions A and C in Fig.\ \ref{fig:modes_operation}
reflects the fact that kinesin is a unidirectional motor.

The response and fluctuations of a motor are quantified,
respectively, by a response matrix $\lambda_{ij}$ and by a
diffusion matrix: $ 2 D_{ij}=  {\partial z_i \partial
z_j}\,\vartheta [\,0,0 ]$. The physical meanings of $\lambda_{ij}$
are: $\lambda_{11} \equiv \partial \hat{v} /\partial f$  is the
mobility, $\lambda_{22} \equiv \partial r /\partial \Delta \mu$ is
the chemical admittance, and more importantly, $\lambda_{12}
\equiv \partial \hat{v} /\partial \Delta \mu$ and $\lambda_{21}
\equiv \partial r/\partial f$ are the Onsager coefficients that
quantify the mechanochemical couplings of the motor.
Differentiating Eq.\ (\ref{FT}), we can write:\bb
\begin{array}{ll} \hat{v} \equiv -  \partial_{z_1}\vartheta
[\,0,0] & =
\partial_{z_1}\vartheta [\,f, \Delta
\mu], \\
r \equiv - \partial_{z_2} \vartheta [\,0,0] & = \partial_{z_2}
\vartheta [\,f,\mu]. \label{v and r}
\end{array}
\en
Near equilibrium, where $f$ and $\Delta \mu$ are small, a
Taylor expansion of Eq.\ (\ref{v and r}) leads to $\hat{v}  =
\lambda_{11}^0\,f + \lambda_{12}^0\,\Delta \mu$ and $r =
\lambda_{21}^0\,f + \lambda_{22}^0\,\Delta \mu$, with
$\lambda_{ij}^0 =
\partial_{z_i}\partial_{z_j} \vartheta [\,0,0] \equiv D_{ij}$,
which are the Einstein relations, and $\lambda_{12}^0 \equiv
\partial_{z_2}\partial_{z_1} \vartheta [\,0,0] =  \partial_{z_1}
\partial_{z_2}\vartheta [\,0,0] \equiv \lambda_{21}^0$, which is
the Onsager relation.  Thus, FT captures the response and
fluctuations near equilibrium \cite{gallavotti,gaspard}.

Away from equilibrium, we expect that Onsager and Einstein
relations are no longer valid. To quantify their violations, we
introduce $\Delta \lambda \equiv \lambda_{12}- \lambda_{21}$ and
four ``temperature"-like quantities, $T_{ij} \equiv
D_{ij}/\lambda_{ij}$. Of course, these effective temperatures are
not thermodynamic temperatures: they are merely one of the ways to quantify
deviations of Einstein relations. Via FT, we obtain the following general
characterizations for these quantities. First,
at sufficiently small driving, we find $2 \Delta \lambda \approx
(\partial_{\Delta \mu} \vartheta_{11} -\partial_f \vartheta_{12})
f +(\partial_{\Delta \mu} \vartheta_{12} -\partial_f
\vartheta_{22}) \Delta \mu$, where $\vartheta_{ij} \equiv
{\partial z_i \partial z_j}\,\vartheta [\,f/2,\Delta \mu/2 ]$. In
particular, for $f \ll 1$, $\Delta \lambda \propto \Delta \mu$.
Thus, active processes in the mechanochemical transduction
mechanism breaks Onsager symmetry and these processes can be
studied via $\Delta \lambda$. Secondly, along $\hat{v}(f,\Delta
\mu) =0$, we find that Eq.\ (\ref{FT}) has a special relation:
$\vartheta[z_1,0\,] = \vartheta[\,\delta f -z_1,0\,]$, where
$\delta f = f - f_{\mbox{\scriptsize st}}(\Delta \mu)$. Therefore,
one of the Einstein relations, $\lambda_{11} = D_{11}$, holds near
stalling, since FT implies that $2\,\hat{v} =
\partial^2_{z_1}\vartheta[\,0,0\,] \, \delta f $ for small $\delta
f$. Note that this particular Einstein relation also holds for ratchet models under
similar consitions \cite{sakaguchi}. By the same token, near $r =0$, FT implies that an Einstein
relation holds for $y$, {\em i.e.} $D_{22} = \lambda_{22}$.

\begin{figure}
\includegraphics[height=3.4in]{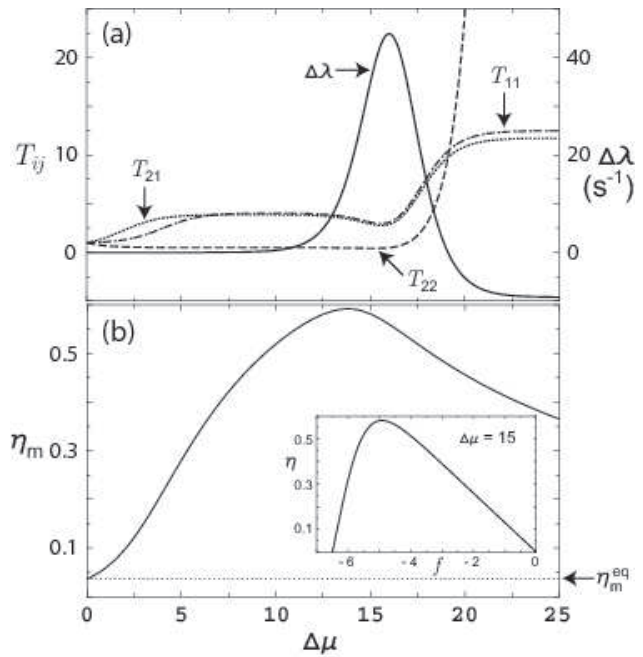}
\caption{(a) Plots of $T_{11}$ (dot-dashed), $T_{21}$ (dotted), $T_{22}$ (dashed), and
$\Delta \lambda$ (solid) vs.\ $\Delta \mu$ in Region A of Fig.\ \ref{fig:modes_operation} with
small $f$.  $T_{ij}$ characterize the fluctuation-response ratios (see text),
and $\Delta \lambda$ quantifies the breaking of Onsager symmetry.
(b) Local maximum of the efficiency $\eta_{\,\mbox{\scriptsize m}}$ vs. $\Delta \mu$.
Note that $\eta_{\,\mbox{\scriptsize m}}$ is substantially larger than
$\eta_{\mbox{\scriptsize m}}^{\mbox{\scriptsize eq}}$ (the dotted line). Note also that
the absolute maximum, which occurs at about $\Delta \mu \approx 15$,
roughly corresponds also to the maximum of $\Delta \lambda$. Inset: Efficiency vs.\
normalized force for $\Delta \mu=15$. The parameters
used in both (a) and (b) are the same as those used to generate Fig.\ \ref{fig:modes_operation}.}
\label{fig:effective temp}
\end{figure}

For our two-state model, we can fully investigate the behaviors of
$\Delta \lambda$ and $T_{ij}$. Let us focus on region A of
Fig.~\ref{fig:modes_operation} and $-f \ll 1$, so that
$\lambda_{ij}$ and $T_{ij}$ depend only on $\Delta \mu$. In Fig.\
\ref{fig:effective temp}a, we display $\Delta \lambda$ and the
three distinct $T_{ij}$ (see below) as a function of $\Delta \mu$.
We observe that for small $\Delta \mu$, $\Delta \lambda$ rises
linearly with $\Delta \mu$, in agreement with the FT prediction,
and that for larger $\Delta \mu$, $\Delta \lambda$ exhibits a
maximum. Moreover, for large $\Delta \mu$, we find that $\Delta
\lambda$ approaches to a constant value. The latter observation
can be understood from a simple argument. When $\Delta \mu \gg 1$,
the transitions between the states $a$ and $b$ of the motor are
limited by the $\beta$-pathways. Therefore, we can write $r \simeq
\omega\,e^{-\theta^{-}_b f} + \omega'\,e^{\theta^{+}_b f}$, which
implies that for small $f$, $\Delta \lambda \simeq
\omega\,\theta^{-}_b -\omega'\,\theta^{+}_b$ since $\lambda_{12}
\approx 0$ for large $\Delta \mu$. Thus, the underlying asymmetry
of the substrate determines the degree by which the Onsager
symmetry is broken.

In Region A and $-f \ll 1$, the $T_{ij}$ also exhibit interesting
behaviors. First, we note that the run length $\ell$ - the
distance moved per ATP hydrolyzed - is independent of $\Delta
\mu$: $\ell \equiv v/r = 2d\,(\alpha\,\omega' - \alpha' \omega ) /
[ (\alpha+ \alpha' )(\omega +\omega')] < 2d$. With the help of FT,
we find that $T_{12}= T_{22}$ for any $\Delta \mu$.  Therefore,
there are only three ``effective" temperatures instead of four, as
one might naturally suppose. As shown in Fig.\ \ref{fig:effective
temp}a, we observe that all distinct $T_{ij}$ start off at $T_{ij}
=1$ near equilibrium, as expected, and for large $\Delta \mu$,
$T_{22} \sim e^{\Delta \mu}$ diverges exponentially whereas
$T_{21}$ and $T_{11}$ approach to finite values. Secondly, for
tightly coupled motors, $\ell/(2d) \sim 1$, we find that $T_{11}$
is nearly identical to $T_{21}$ (see Fig.\ \ref{fig:effective
temp}a). Therefore, in this case, only two ``effective"
temperatures characterize motors' fluctuations.

In addition, our framework allows us to investigate the
thermodynamic efficiency, an important quantity that also
characterizes the working of a motor \cite{ken}.  In region A, it
is defined as the ratio of the work performed to the chemical
energy input: $\eta \equiv -f \hat{v}/(r \Delta \mu)$
\cite{armand1}. By definition, $\eta$ vanishes at $f=0$ and at the
stalling force $f_{\mbox{\scriptsize st}}$. Therefore, it has a
local maximum $\eta_{\mbox{\scriptsize m}}(\Delta \mu)$ for some
$f_{\mbox{\scriptsize m}}(\Delta \mu)$ between
$f_{\mbox{\scriptsize st}} < f_{\mbox{\scriptsize m}} < 0$ (see
Fig.\ \ref{fig:effective temp}b inset).  Near equilibrium,
$\eta_{\mbox{\scriptsize m}}(\Delta \mu)$ has a constant value,
$\eta_{\mbox{\scriptsize m}}^{\mbox{\scriptsize eq}}$, along a
straight line $f_{\mbox{\scriptsize m}}(\Delta \mu)  \propto
\Delta \mu$ inside region A \cite{armand1}. Far from equilibrium,
we find that $\eta_{\mbox{\scriptsize m}}$ has an absolute maximum at
some $\Delta \mu > 1$, and $\eta_{\mbox{\scriptsize m}}$ is
substantially larger than $\eta_{\mbox{\scriptsize
m}}^{\mbox{\scriptsize eq}}$ as shown in Fig.\ \ref{fig:effective
temp}b.  Hence, a motor achieves a higher efficiency in the
far-from-equilibrium regimes \cite{parmeggiani}.

\begin{figure}
\includegraphics[height=1.9in]{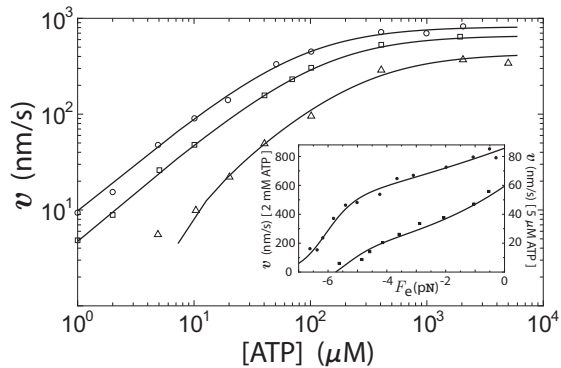}
\caption{Kinesin velocity vs.\ ATP concentration under an external force.
The solid curves are the fits of our model to data from Ref.\ \cite{block}.
From the top down, the plots are for $F_e = -1.05, -3.59$, and $-5.63\,\mbox{pN}$, respectively.
Inset: Kinesin velocity vs.\ force under a fixed ATP concentration.  The solid curves are
fits to the data of Ref.\ \cite{block}. From the top down, the plots are
for $[\mbox{ATP}] = 2\,\mbox{mM}$ and $5\,\mbox{$\mu$M}$, respectively.  }
\label{fig:fits}
\end{figure}

Finally, to discuss the relevance of our framework to kinesin, we
carried out a global fit of kinesin velocity data of
Ref.~\cite{block} to our model at different external forces and
two curves of force vs.\ velocity at different ATP concentrations
(see Fig.\ \ref{fig:fits}).  Assuming that $e^{\Delta \mu} =
k_0\,[\mbox{ATP}]$, we obtain the best-fit values for the
parameters: $\epsilon =  10.81$, $k_0 = 1.4 \cdot
10^{5}\,\mbox{$\mu$M}^{-1}$, $\alpha = 0.57\,\mbox{s}^{-1}$,
$\alpha' = 1.3 \cdot 10^{-6}\,\mbox{s}^{-1}$, $\omega =
3.5\,\mbox{s}^{-1}$, $\omega'= 108.15\,\mbox{s}^{-1}$,
$\theta^{+}_a = 0.25$, $\theta^{-}_a = 1.83$, $\theta^{+}_b=0.08$,
and $\theta^{-}_b = -0.16$.  These values are reasonable within
the accepted biophysical picture of kinesin \cite{howard}. First,
$\epsilon$ and $k_0^{-1}$ represent the typical binding energy
($\sim 10\,k_B T$) of kinesin with microtubules and the ATP
concentration at equilibrium ($\sim 10^{-5}\,\mbox{$\mu$M}$),
respectively. Secondly, $\theta^{-}_a = 1.83$ indicates that the
back-steps (transitions $a \rightarrow b$) of kinesin contain most
of the displacement sensitivity \cite{howard}. Moreover, our
framework allows us to estimate a maximum stalling force of
$-7\,\mbox{pN}$, and more importantly, a run length of $\ell
\simeq 0.97 (2d)$ and a global ATP consumption rate of $r \simeq
111\,\mbox{s}^{-1}$, all in excellent agreement with known values
\cite{howard}. Using the above parameters, we constructed the
diagram of operation for kinesin (Fig.~\ref{fig:modes_operation}),
we made predictions about $\Delta \lambda$ and $T_{ij}$
(Fig.~\ref{fig:effective temp}a), and we obtained the efficiency
for kinesin (Fig.\ \ref{fig:effective temp}b). In particular, we
find that $T_{11} \sim 10\,T,$ the maximum value of $\Delta
\lambda \sim 45\,\mbox{pN}^{-1}\mbox{s}^{-1}$, and $\Delta \lambda
\sim -10\,\mbox{pN}^{-1}\mbox{s}^{-1}$ at large $\Delta \mu$.
Under typical physiological conditions ($\Delta \widetilde{\mu} \sim 10 -
25\,\mbox{$k_B T$}$), kinesin operates at an efficiency in the range of $40 -
60\%$, also in agreement with experiments \cite{howard}. Lastly,
we point out a remarkable feature: the absolute maximum of
$\eta_{\mbox{\scriptsize m}}$ occurs approximately at a $\Delta
\mu $ at which $\Delta \lambda$ is also a maximum, corresponding
to an energy scale of $15 - 20\,k_B T$ (see Fig.\
\ref{fig:effective temp}). It is interesting to note that
kinesins operate most efficiently in an energy scale corresponding
to the energy available from ATP hydrolysis.

In conclusion, FT links a set of physical quantities that reveal
 the mechanochemical couplings
of a motor and our results support a growing consensus that FT
provides a possible organizing
principle for driven active systems.

We acknowledge important discussions with A.\ Ajdari, Y.\
Hatwalne, J.F.\ Joanny, F. J\"{u}licher, T.C. Lubensky, and J.\
Prost. We acknowledge support from the ESPCI (for A.W.C.L.) and
from a CEFIPRA grant (for D.L.).

\end{document}